\documentclass[12pt]{iopart}
\usepackage{graphicx}
\begin{document}
\title{A New Two-Parameter Family of Potentials with a Tunable Ground State}
\author{Jonathan M Fellows and Robert A Smith}
\address{School of Physics and Astronomy, University of Birmingham, Edgbaston, Birmingham B15 2TT, United Kingdom}
\eads{\mailto{fellowsjm@theory.bham.ac.uk}, \mailto{ras@th.ph.bham.ac.uk}}
\begin{abstract}
In a previous paper \cite{FS2009} we solved a countably infinite family of one-dimensional Schr\"odinger equations by showing that they were
supersymmetric partner potentials of the standard quantum harmonic oscillator. In this work we extend these results to find the
complete set of real partner potentials of the harmonic oscillator, showing that these depend upon two continuous parameters. Their
spectra are identical to that of the harmonic oscillator, except that the ground state energy becomes a tunable parameter. We finally
use these potentials to analyse the physical problem of Bose-Einstein condensation in an atomic gas trapped in a dimple potential.
\end{abstract}

\section{Introduction}

In a previous paper in this journal \cite{FS2009}, we found the exact solution of the Schr\"odinger equation for a family of potentials consisting of a harmonic oscillator term plus a rational function. The Schr\"odinger equation for these potentials is
\begin{equation*}
-{d^2\psi\over dx^2}+\left[x^2+2{{\cal H}_p'(x)^2-{\cal H}_p(x){\cal H}_p''(x)\over{\cal H}_p(x)^2}\right]\psi=E\psi,
\end{equation*}
where the ${\cal H}_p(x)$ are the pseudo-Hermite polynomials defined by
\begin{equation*}
{\cal H}_p(x)=(-i)^pH_p(ix)=e^{-x^2}{d^p\over dx^p}\left[e^{x^2}\right].
\end{equation*}
These pseudo-Hermite polynomials, ${\cal H}_p(x)$, are just the regular Hermite polynomials, $H_p(x)$, with all the minus signs removed. For even $p$ the potentials take the form of a harmonic well with an additional attractive dimple of depth $4p$ at the origin. The bound states have energies, $E_k=2k+3$, and wavefunctions,
\begin{equation*}
\psi_k(x)=\left[{1\over 2^{k+1}k!(k+p+1)\sqrt{\pi}}\right]^{1\over 2}
\left[H_{k+1}(x)+{{\cal H}_p'(x)\over{\cal H}_p(x)}H_k(x)\right]e^{-x^2/2},
\end{equation*}
where $k=0,1,2,3,\dots$ In addition, there is a new ground state with energy, $E_{-1}=1-2p$, and wavefunction,
\begin{equation*}
\psi_{-1}(x)=\left[{2^{p/2}p!\over\sqrt{\pi}}\right]^{1/2}{e^{-x^2/2}\over{\cal H}_p(x)}.
\end{equation*}
It is as if the attractive dimple of depth $4p$ has taken the ground state energy level of the harmonic oscillator and lowered it by an amount $2p$, whilst leaving all the other energy levels fixed. In the above solution, $p$ has to be an even integer, but one could imagine a situation where
$p$ is an arbitrary positive real number. This would yield a potential with a spectrum consisting of an evenly spaced ladder with a ground state energy level which may be lowered at will by the addition of an attractive dimple. Such a potential, as well as being interesting from a purely mathematical point of view, may have practical application in the area of atomic Bose-Einstein condensation. It has been established, both theoretically \cite{PS2003} 
and experimentally \cite{Stamper1998}, that the adiabatic addition of a narrow attractive potential to a harmonic trap can lead to Bose-Einstein condensation at a higher critical temperature. The theoretical justification for this effect is that an additional bound state may be formed in the superposed potential, which may have energy as low as the chemical potential of the unperturbed system, leading to macroscopic occupation of that level. The problem with such a heuristic explanation is that when two potentials are superposed, one does not usually obtain a spectrum which bears any resemblance to that of either of the superposed potentials. Our potential is therefore special in that the effect of adding an attractive dimple
of the appropriate form has the sole effect of lowering its ground state energy. This makes it a natural starting point for future theoretical work on the dimple problem in atom traps.

In this paper we generalize our previous family of soluble potentials to allow the parameter $p$ to take non-integer values, and thus obtain an energy spectrum with a continuously tunable ground state. Since our previous potentials were solved by noting that they are supersymmetric partner potentials of the harmonic oscillator, we proceed by determining the most general potential which is a partner of the harmonic oscillator. This leads to a two-parameter family of soluble potentials, of which the desired potentials are a subset.

The rest of the paper is organized as follows. In section 2 we give a prescription for generating a two-parameter family of soluble potentials from a known soluble potential using the factorization approach. In section 3 we apply this prescription to obtain the set of all potentials which are partners of the harmonic oscillator, and derive the eigenvalues and eigenfunctions for these potentials. In section 4 we plot a representative sample
of these potentials and their eigenfunctions. In section 5 we analyze the problem of Bose-Einstein condensation in one of the dimple potentials
derived in section 3, and in section 6 we make our summary and conclusions.

\section{The factorization approach}

Consider the one-dimensional single-particle Schr\"odinger equation,
\begin{equation*}
H_1\psi(x)=\left[-{d^2\over dx^2}+V_1(x)\right]\psi(x)=E\psi(x).
\end{equation*}
The idea of the factorization approach (see \cite{Cooper2001} and references therein) is to write the Hamiltonian operator, $H_1$, as a
product of two first-order differential operators,
\begin{equation*}
H_1=A^{\dagger}A,
\end{equation*}
where
\begin{equation*}
A={d\over dx}+W(x),\qquad
A^{\dagger}=-{d\over dx}+W(x),
\end{equation*}
so that
\begin{equation*}
V_1(x)=W^2(x)-W'(x).
\end{equation*}
If we now define the operator $H_2=AA^{\dagger}$ by reversing the order of $A$ and $A^{\dagger}$, we see that $H_2$ is a Hamiltonian operator corresponding to a new potential, $V_2(x)$,
\begin{equation*}
H_2=AA^{\dagger}=-{d^2\over dx^2}+V_2(x),\qquad
V_2(x)=W(x)^2+W'(x).
\end{equation*}
The potentials $V_1(x)$ and $V_2(x)$ are known as supersymmetric partner potentials.

The eigenvalues and eigenfunctions of $H_1$ and $H_2$ are related. If $\psi^{(1)}_n$ is an eigenfunction of $H_1$ with energy eigenvalue $E^{(1)}_n$, then $A\psi^{(1)}_n$ is an eigenfunction of $H_2$ with energy eigenvalue $E^{(1)}_n$ since
\begin{equation*}
H_2\left[A\psi^{(1)}_n\right]=
AA^{\dagger}A\psi^{(1)}_n=
A\left[H_1\psi^{(1)}_n\right]=
A\left[E^{(1)}_n\psi^{(1)}_n\right]=
E^{(1)}_n\left[A\psi^{(1)}_n\right].
\end{equation*}
Similarly if $\psi^{(2)}_n$ is an eigenfunction of $H_2$ with eigenvalue $E^{(2)}_n$, then $A^{\dagger}\psi^{(2)}_n$ is an eigenfunction of $H_1$ with energy eigenvalue $E^{(2)}_n$ since
\begin{equation*}
H_1\left[A^{\dagger}\psi^{(2)}_n\right]=
A^{\dagger}AA^{\dagger}\psi^{(2)}_n=
A^{\dagger}\left[H_2\psi^{(2)}_n\right]=
A^{\dagger}\left[E^{(2)}_n\psi^{(2)}_n\right]=
E^{(2)}_n\left[A^{\dagger}\psi^{(2)}_n\right].
\end{equation*}
It follows that if we know how to exactly solve one of $H_1$ or $H_2$, we can immediately derive the exact solution of the other.

Suppose that we know the exact solution of the Hamiltonian $H_2$. Let us find its complete set of partner Hamiltonians, $H_1$. The function $W(x)$ in the factorization must satisfy the equation
\begin{equation*}
W(x)^2+W'(x)=V_2(x)-\lambda,
\end{equation*}
where we have noted that the addition of an arbitrary constant $\lambda$ does not affect the solution of $H_2$. To solve this equation we make the standard substitution
\begin{equation*}
W(x)={d\over dx}\ln{\phi(x)}={\phi'(x)\over\phi(x)},
\end{equation*}
which yields the result
\begin{equation*}
-{d^2\phi\over dx^2}+V_2(x)\phi(x)=\lambda\phi(x).
\end{equation*}
It follows that $\phi(x)$ is a solution of the original Schr\"odinger equation for the known soluble potential $V_2(x)$. The central idea of our approach is that $\phi(x)$, whilst a solution of the Schr\"odinger differential equation, does not have to be a normalizable wavefunction. There is therefore a general solution, $\phi_\lambda(x)$, for any value of the pseudo-energy parameter, $\lambda$,
\begin{equation*}
\phi_\lambda(x)=C\phi_{\lambda,1}(x)+D\phi_{\lambda,2}(x),
\end{equation*}
where $\phi_{\lambda,1}(x)$ and $\phi_{\lambda,2}(x)$ are two independent solutions of the Schr\"odinger equation. If $V_2(x)$ is even, these solutions will have definite parity, and without loss of generality we can assume $\phi_{\lambda,1}(x)$ is even and $\phi_{\lambda,2}(x)$ is odd. The function $W(x)$ is then given by
\begin{equation*}
W_{\lambda,s}(x)={\phi_{\lambda,1}'(x)+s\phi_{\lambda,2}'(x)\over \phi_{\lambda,1}(x)+s\phi_{\lambda,2}(x)},
\end{equation*}
where $s=D/C$. We see that $W_{\lambda,s}(x)$ depends upon the two real parameters $\lambda$ and $s$, and therefore so does $V_1(x)$. The only restriction we will impose on $\lambda$ and $s$ is that the function $\phi_\lambda(x)$ has no zeros, so that the potential $V_1(x)$ is non-singular. This means that $\lambda$ must be less than or equal to the ground state energy of $H_2$.

\section{The complete set of partner potentials of the harmonic oscillator}

In this section we will take the harmonic potential, $V_2(x)=x^2$, to be the known soluble potential,
and derive the complete set of partner potentials, $V_1(x)$.

\subsection{Schr\"odinger equation for harmonic oscillator}

To solve the Schr\"odinger equation for the harmonic oscillator,
\begin{equation*}
-{d^2\phi\over dx^2}+x^2\phi=\lambda\phi,
\end{equation*}
we first remove the asymptotic behaviour by setting
\begin{equation*}
\phi(x)=\xi(x)e^{-x^2/2},
\end{equation*}
which yields the equation
\begin{equation*}
{d^2\xi\over dx^2}-2x{d\xi\over dx}+(\lambda-1)\xi=0.
\end{equation*}
If we change the independent variable to $z=x^2$, the latter equation becomes,
\begin{equation*}
z{d^2\xi\over dz^2}+({\textstyle{1\over 2}}-z){d\xi\over dz}+{\textstyle{1\over 4}}(\lambda-1)\xi=0,
\end{equation*}
which has the form of the confluent hypergeometric equation \cite{GnR},
\begin{equation*}
z{d^2\xi\over dz^2}+(c-z){d\xi\over dz}-a\xi=0,
\end{equation*}
with $a={1\over 4}(1-\lambda)$ and $c={1\over 2}$. This can be solved using the method of Frobenius to give the two solutions
\begin{equation*}
\Phi(a,c;z)\quad\hbox{and}\quad z^{1-c}\Phi(a+1-c,2-c;z),
\end{equation*}
where
\begin{equation*}
\Phi(a,b;z)=\sum_{n=0}^\infty {\Gamma(a+n)\Gamma(c)\over\Gamma(a)\Gamma(c+n)}{z^n\over n!}.
\end{equation*}
It follows that the two independent solutions of the harmonic oscillator Schr\"odinger equation are
\begin{equation*}
\Phi({\textstyle{1-\lambda\over 4}},{\textstyle{1\over 2}};x^2)e^{-x^2/2}\qquad\hbox{and}\qquad
x\Phi({\textstyle{3-\lambda\over 4}},{\textstyle{3\over 2}};x^2)e^{-x^2/2}.
\end{equation*}
To obtain the standard harmonic oscillator eigenfunctions, $\psi^{(2)}_k(x)$, we write $\lambda=2k+1$ and note that the confluent hypergeometric function, $\Phi(a,c;z)$, becomes a terminating polynomial if parameter $a$ is a negative integer. If $k$ is even, then the even solution
terminates to give
\begin{equation*}
\psi^{(2)}_k(x)=\Phi({\textstyle-{k\over 2}},{\textstyle{1\over 2}};x^2)e^{-x^2/2},
\end{equation*}
whilst if $k$ is odd, then the odd solution terminates to give
\begin{equation*}
\psi^{(2)}_k(x)=x\Phi({\textstyle{1-k\over 2}},{\textstyle{3\over 2}};x^2)e^{-x^2/2}.
\end{equation*}
These can be written in terms of Hermite polynomials using the identities \cite{GnR},
\begin{eqnarray*}
H_{2n}(x)&=(-1)^n {(2n)!\over n!} \Phi(-n,{\textstyle{1\over 2}};x^2)\\
H_{2n+1}(x)&=(-1)^n 2 {(2n+1)!\over n!} x \Phi(-n,{\textstyle{3\over 2}};x^2),
\end{eqnarray*}
so that the normalized solutions of the harmonic oscillator are
\begin{equation*}
\psi^{(2)}_k(x)=\left[{1\over 2^k k! \sqrt{\pi}}\right]^{1/2}
H_k(x)e^{-x^2/2}.
\end{equation*}

In our previous paper, we found suitable nodeless solutions, $\phi(x)$, by employing the internal symmetry $x\rightarrow ix$,
$\lambda\rightarrow -\lambda$ in the harmonic oscillator Schr\"odinger equation. Our two general solutions then become
\begin{equation*}
\phi_{p,1}(x)=\Phi({\textstyle-{p\over 2}},{\textstyle{1\over 2}};-x^2)e^{x^2/2},\qquad
\phi_{p,2}(x)=x\Phi({\textstyle{1-p\over 2}},{\textstyle{3\over 2}};-x^2)e^{x^2/2};
\end{equation*}
if $p$ is an even or odd integer then $\phi_{p,1}(x)$ or $\phi_{p,2}(x)$, respectively, are the solutions employed in our previous
paper. The point of what we are doing here is to ensure that our notation aligns with that of our previous paper for integer $p$;
however we will henceforth be considering the general case where $p$ may be non-integer. Note that the confluent hypergeometric
identity \cite{GnR},
\begin{equation*}
\Phi(a,c;-z)=e^{-z}\Phi(c-a,c;z),
\end{equation*}
implies that
\begin{equation*}
\phi_{p,1}(x)=\Phi({\textstyle{p+1\over 2}},{\textstyle{1\over 2}};x^2)e^{-x^2/2},\qquad
\phi_{p,2}(x)=x\Phi({\textstyle{p\over 2}+1},{\textstyle{1\over 2}};x^2)e^{-x^2/2},
\end{equation*}
which we recognize as the general solutions for $\lambda=-(2p+1)$, as we would expect.

\subsection{Even parity partner potentials}

To construct new soluble potentials, $V_1(x)$, we first evaluate the superpotential $W(x)$. Let us first
restrict ourselves to the even potentials by using the solution $\phi_{p,1}(x)$ so that
\begin{eqnarray*}
W(x)&={d\over dx}\ln\phi_{p,1}(x)\\
&=-x+{d\over dx}\ln\Phi({\textstyle{p+1\over 2}},{\textstyle{1\over 2}};x^2)\\
&=-x+2(p+1)x{\Phi({\textstyle{p+3\over 2}},{\textstyle{3\over 2}};x^2)\over
\Phi({\textstyle{p+1\over 2}},{\textstyle{1\over 2}};x^2)},
\end{eqnarray*}
where we used the identity \cite{GnR},
\begin{equation*}
{d\over dz}\Phi(a,c,z)={a\over c}\Phi(a+1,c+1,z).
\end{equation*}
From this we generate
\begin{equation*}
V_2(x)=W^2(x)+W'(x)=x^2+2p+1=V_H(x)+2p+1,
\end{equation*}
which is clearly the harmonic oscillator potential, $V_H(x)$, with the offset $2p+1$. We will choose the offset
for $V_1(x)$ so that it asymptotically approaches $V_H(x)$ at infinity. This leads to an offset of $2p-1$. Hence
\begin{equation*}
V_1(x)=W^2(x)-W'(x)=2W^2(x)-V_1(x)=V_p(x)+2p-1,
\end{equation*}
giving the final result for the partner potential
\begin{equation*}
V_p(x)=x^2-4xw(x)+2w^2(x)-4p,
\end{equation*}
where we define for convenience
\begin{equation*}
w(x)=2(p+1)x{\Phi({\textstyle{p+3\over 2}},{\textstyle{3\over 2}};x^2)\over
\Phi({\textstyle{p+1\over 2}},{\textstyle{1\over 2}};x^2)}.
\end{equation*}
The normalized eigenfunctions of this potentials are then given by
\begin{eqnarray*}
\psi^{p}_k(x)&=\left[{1\over E_k^{(2)}}\right]^{1/2}A^{\dag}\psi_k^{(2)}(x)\\
&=\left[{1\over 2^{k+1}k!(k+p+1)\sqrt{\pi}}\right]^{1/2}
\left[-{d\over dx}-x+w(x)\right]H_k(x)e^{-x^2/2}\\
&=\left[{1\over 2^{k+1}k!(k+p+1)\sqrt{\pi}}\right]^{1/2}
\Big[w(x)H_k(x)-2kH_{k-1}(x)\Big]e^{-x^2/2},
\end{eqnarray*}
with energy $E_k^p=2k+3$ for $k=0,1,2,3,\dots$. The new ground state is given by
\begin{equation*}
\psi_{-1}^p(x)\propto {1\over\phi_{1,p}(x)}=
{e^{x^2/2}\over\Phi({\textstyle{p+1\over 2}},{\textstyle{1\over 2}};x^2)}.
\end{equation*}
The normalization factor can be guessed by analytically continuing from the result
for even integer $p$ derived in the previous paper \cite{FS2009},
\begin{equation*}
\psi^p_{-1}(x)=\left[{p!\,2^p\over\sqrt{\pi}}\right]^{1/2}{e^{-x^2/2}\over{\cal H}_p(x)},
\end{equation*}
and using the result that
\begin{equation*}
{\cal H}_p(x)={\Gamma({\textstyle{p\over 2}+1})\over\Gamma(p+1)}
\Phi({\textstyle{p+1\over 2}},{\textstyle{1\over 2}};x^2).
\end{equation*}
This gives the final result for the normalized ground state wavefunction,
\begin{equation*}
\psi_{-1}^p(x)=\Gamma({\textstyle{p\over 2}+1})
\left[{2^p\over\Gamma(p+1)\sqrt{\pi}}\right]^{1/2}
{e^{x^2/2}\over\Phi({\textstyle{p+1\over 2}},{\textstyle{1\over 2}};x^2)},
\end{equation*}
with energy, $E_{-1}^p=1-2p$. We see that the spectrum of $V_p(x)$ is identical to that
of the harmonic oscillator except that the ground state energy has been lowered by the
amount $2p$.

Notice that in the above we are guessing the normalization integral,
\begin{equation*}
I_{p,0}=\int_{-\infty}^{\infty} {e^{x^2}\over\Big[\Phi({\textstyle{p+1\over 2}},{\textstyle{1\over 2}};x^2)\Big]^2}\,dx
={\sqrt{\pi}\,\Gamma(p+1)\over 2^p\,\Gamma({\textstyle{p+1\over 2}})^2},\qquad\qquad (p>-1)
\end{equation*}
which we previously proved \cite{FS2009} for even integer $p$. A formal proof of this result is given in Appendix A.

\subsection{General (skewed) partner potentials}

Let us now consider the most general non-singular partner potentials, $V_1(x)$. We do this by using the
general solution,
\begin{equation*}
\phi_{p,s}(x)=\phi_{p,1}(x)+s\phi_{p,2}(x)
=\left[\Phi({\textstyle{p+1\over 2}},{\textstyle{1\over 2}};x^2)
+sx\Phi({\textstyle{p\over 2}+1},{\textstyle{3\over 2}};x^2)\right]e^{-x^2/2},
\end{equation*}
where the skewness parameter, $s$, determines how much odd solution we add to the even solution. We choose
$s$ so that $\phi_p(x)$ has no zeros, and hence $V_1(x)$ has no singularities. To determine the maximum
allowed range of $s$, we use the result for the asymptotic limit of the confluent hypergeometric function \cite{GnR},
\begin{equation*}
\Phi(a,c,z)\sim {\Gamma(c)\over \Gamma(c-a)}(-z)^{-a}
+{\Gamma(c)\over\Gamma(a)}e^{z}z^{a-c},\qquad\qquad |z|\rightarrow\infty
\end{equation*}
so that
\begin{equation*}
\phi_{p,s}(x)\sim\left[{\Gamma({\textstyle{1\over 2}})\over\Gamma({\textstyle{p+1\over 2}})}
+s{\Gamma({\textstyle{3\over 2}})\over\Gamma({\textstyle{p\over 2}+1})}\hbox{sgn}(x)\right]
|x|^p e^{x^2/2}.
\end{equation*}
In order for $\phi_{p,s}(x)$ to not change sign, we require the condition
\begin{equation*}
|s|<s_{max}=2{\Gamma({\textstyle{p\over 2}+1})\over\Gamma({\textstyle{p+1\over 2}})}.
\end{equation*}
Note that we need only consider $0\le s\le s_{max}$, since the potentials for negative $s$ are just
the mirror images of those for positive $s$. All the formulae for the eigenvalues and eigenfunctions
are the same as in the even case, except that $w(x)$ is now given by
\begin{equation*}
w(x)={2(p+1)x\Phi({\textstyle{p+3\over 2}},{\textstyle{3\over 2}};x^2)+
sx\Phi({\textstyle{p\over 2}+1},{\textstyle{3\over 2}};x^2)+
{2\over 3}(p+2)sx^2\Phi({\textstyle{p\over 2}+2},{\textstyle{5\over 2}};x^2)\over
\Phi({\textstyle{p+1\over 2}},{\textstyle{1\over 2}};x^2)
+sx\Phi({\textstyle{p\over 2}+1},{\textstyle{3\over 2}};x^2)}.
\end{equation*}
The new ground state is given by
\begin{equation*}
\psi_{-1}^p(x)\propto {1\over\phi_{p,s}(x)}={e^{x^2/2}\over\Phi({\textstyle{p+1\over 2}},{\textstyle{1\over 2}};x^2)
+sx\Phi({\textstyle{p\over 2}+1},{\textstyle{3\over 2}};x^2)}.
\end{equation*}
To normalize this we need to evaluate the normalization integral
\begin{equation*}
I_{p,s}=\int_{-\infty}^{\infty} {e^{x^2}\over\left[\Phi({\textstyle{p+1\over 2}},{\textstyle{1\over 2}};x^2)
+sx\Phi({\textstyle{p\over 2}+1},{\textstyle{3\over 2}};x^2)\right]^2}\,dx.
\end{equation*}
We can guess the value of this integral by noting that we know its value at $s=0$, and it must diverge at
$s=\pm s_{max}$, to obtain a sensible interpolation formula
\begin{equation*}
I_{p,s}={I_{p,0}\over (1-s^2/s^2_{max})}={\sqrt{\pi}\,\Gamma(p+1)\over
2^p\left[\Gamma({\textstyle{p\over 2}+1})^2-{s^2\over 4}\Gamma({\textstyle{p+1\over 2}})^2\right]}.
\end{equation*}
A formal proof of this result is given in Appendix A.

\subsection{Summary}
We have derived the most general non-singular partner potentials of the harmonic
oscillator in terms of two continuous real parameters $p$ and $s$, and have presented their
eigenvalues and eigenfunctions. All these potentials have the same spectrum as the harmonic
oscillator, except that the ground state has been lowered by an amount $2p$. The parameter $s$
is a skewness parameter which changes the shape of the potential without changing its spectrum.
The allowed ranges of the parameters $p$ and $s$ are
\begin{equation*}
p>-1, \qquad
|s|<s_{max}=2{\Gamma({\textstyle{p\over 2}+1})\over\Gamma({\textstyle{p+1\over 2}})}.
\end{equation*}
Note that for $-1<p<0$, the ground state energy is raised and can be brought arbitrarily close
to the first excited energy level; we find in this case that the potential has a positive bump
at the centre, and so is a double well. In the following section we will plot the potentials
and their eigenfunctions for representative ranges of $p$ and $s$.

\section{Representative examples of our new soluble potentials}

In the previous section we derived the complete solution of the Schr\"odinger equation for
a two-parameter family of potentials. We now wish to plot the potentials and their eigenfunctions
for suitable regions of the $(p,s)$ parameter space. For ease of notation let us henceforth rescale
$s\rightarrow s/s_{max}$.

\subsection{Symmetric dimple potentials: $p>0$, $s=0$}

\begin{figure}[h]
\begin{tabular}{cc}
\includegraphics[width=3in]{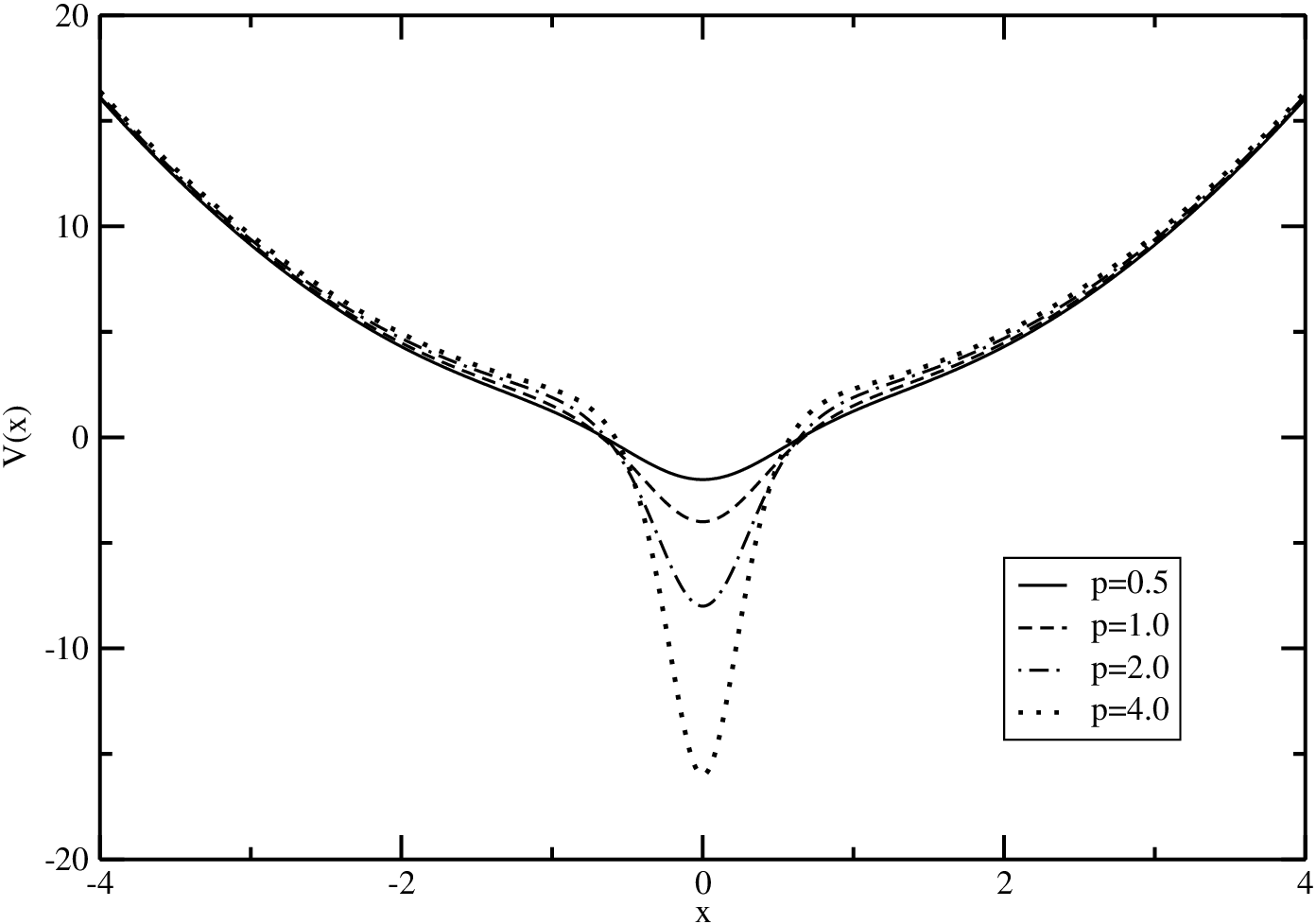}&
\includegraphics[width=3in]{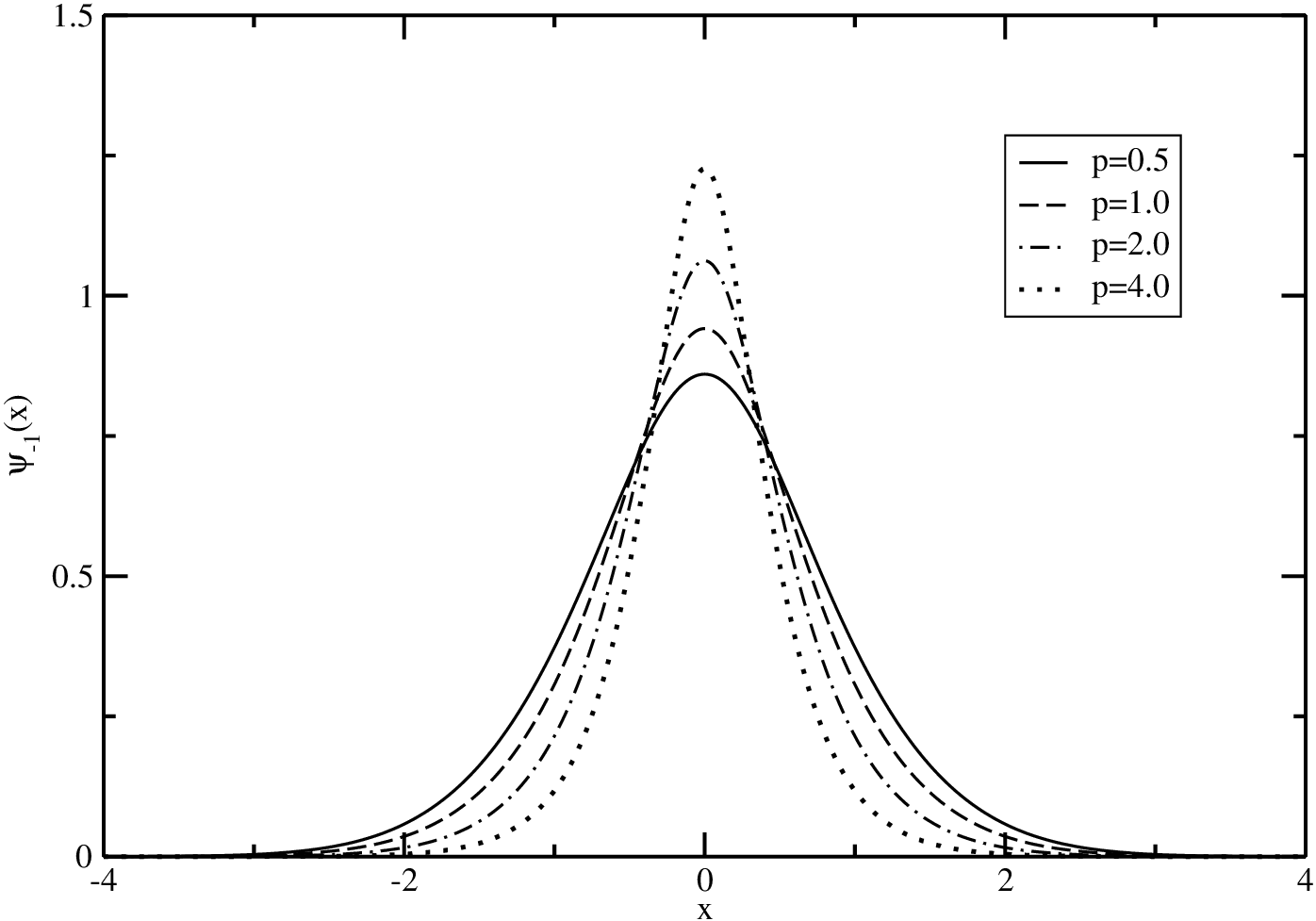}\\
\\
\\
\includegraphics[width=3in]{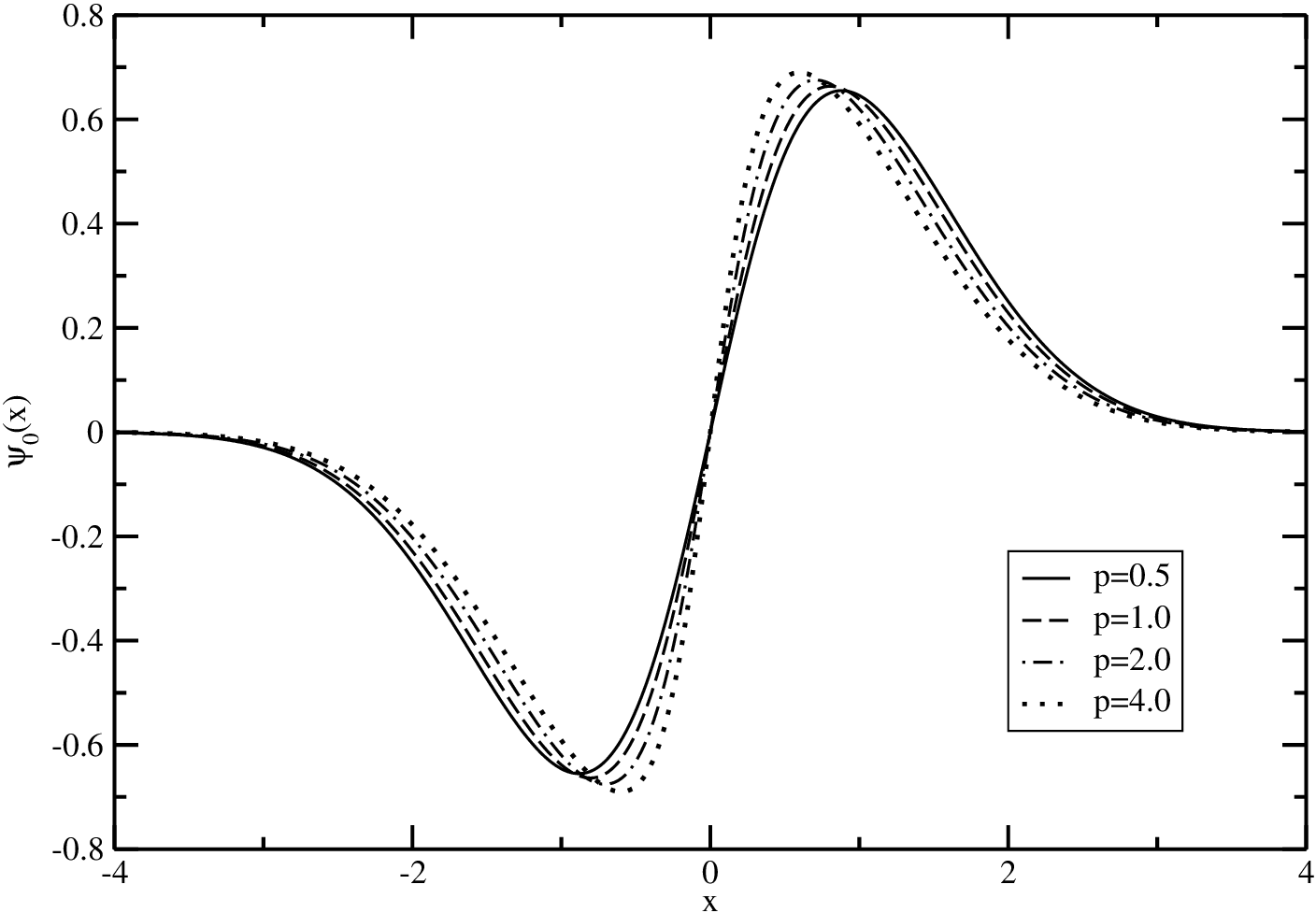}&
\includegraphics[width=3in]{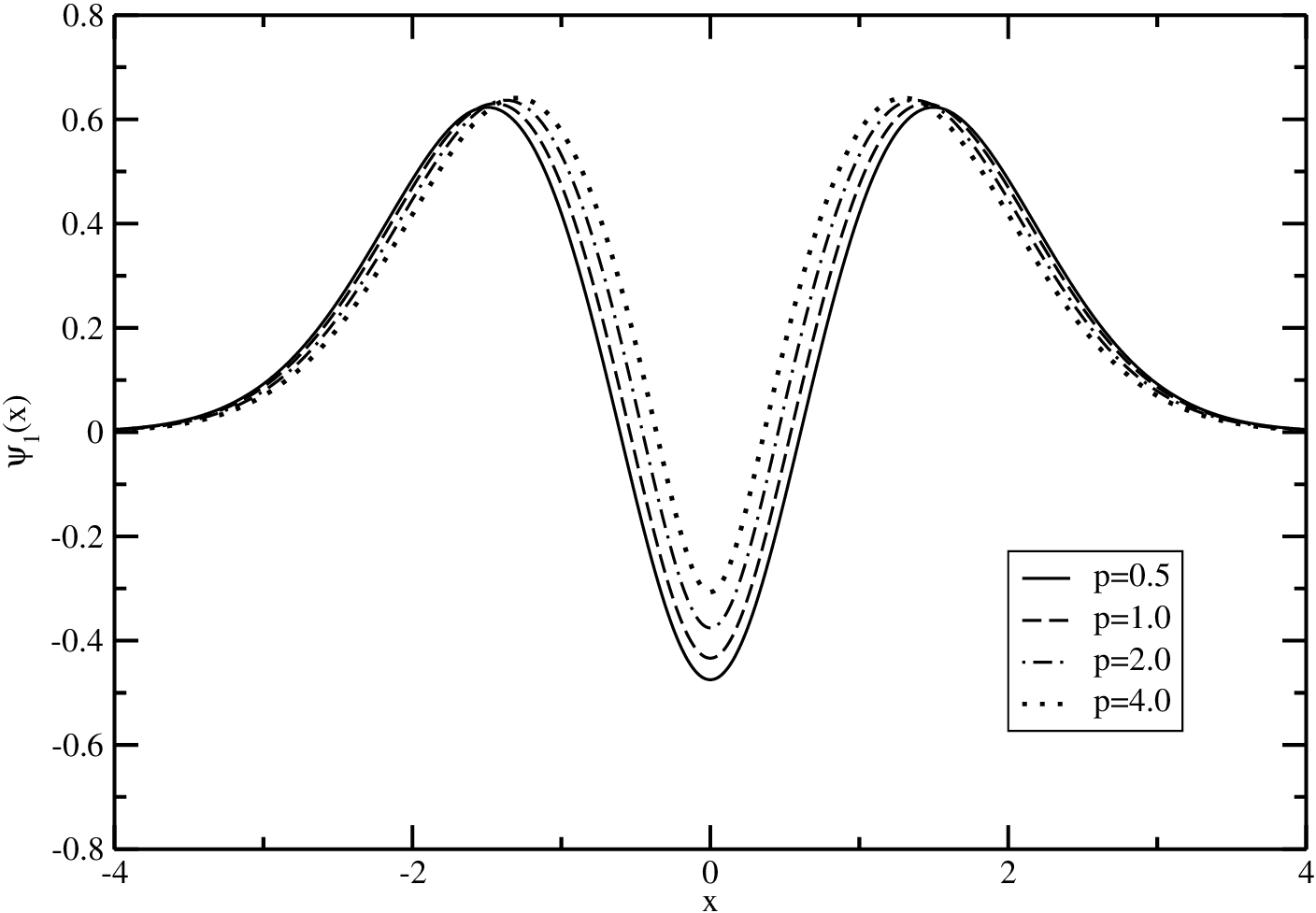}\\
\end{tabular}
\caption{Symmetric dimple potentials and their wavefunctions corresponding to parameter values
$p>0$ and $s=0$: (a) Potentials; (b) Ground state; (c) First excited state; (d) Second excited states.}
\end{figure}

If the skew parameter $s$ equals zero, the partner potential generated will be even. If $p>0$, the
potential will take the form of a harmonic well plus an attractive dimple of depth $4p$ at the origin.
The potentials and their first three wavefunctions are shown in Fig. (1) for $p=0.5$, $1$, $2$ and $4$.
We see that the ground state becomes narrower and taller as the dimple depth increases, whilst the excited
states are less affected. It is as if only the ground state really notices the attractive dimple, and has
its energy pulled down by half the depth of the dimple. The other states keep the same energy and change their
shape only enough to retain orthogonality to the ground state. 

\subsection{Symmetric double well potentials: $0>p>-1$, $s=0$}

\begin{figure}[h]
\begin{tabular}{cc}
\includegraphics[width=3in]{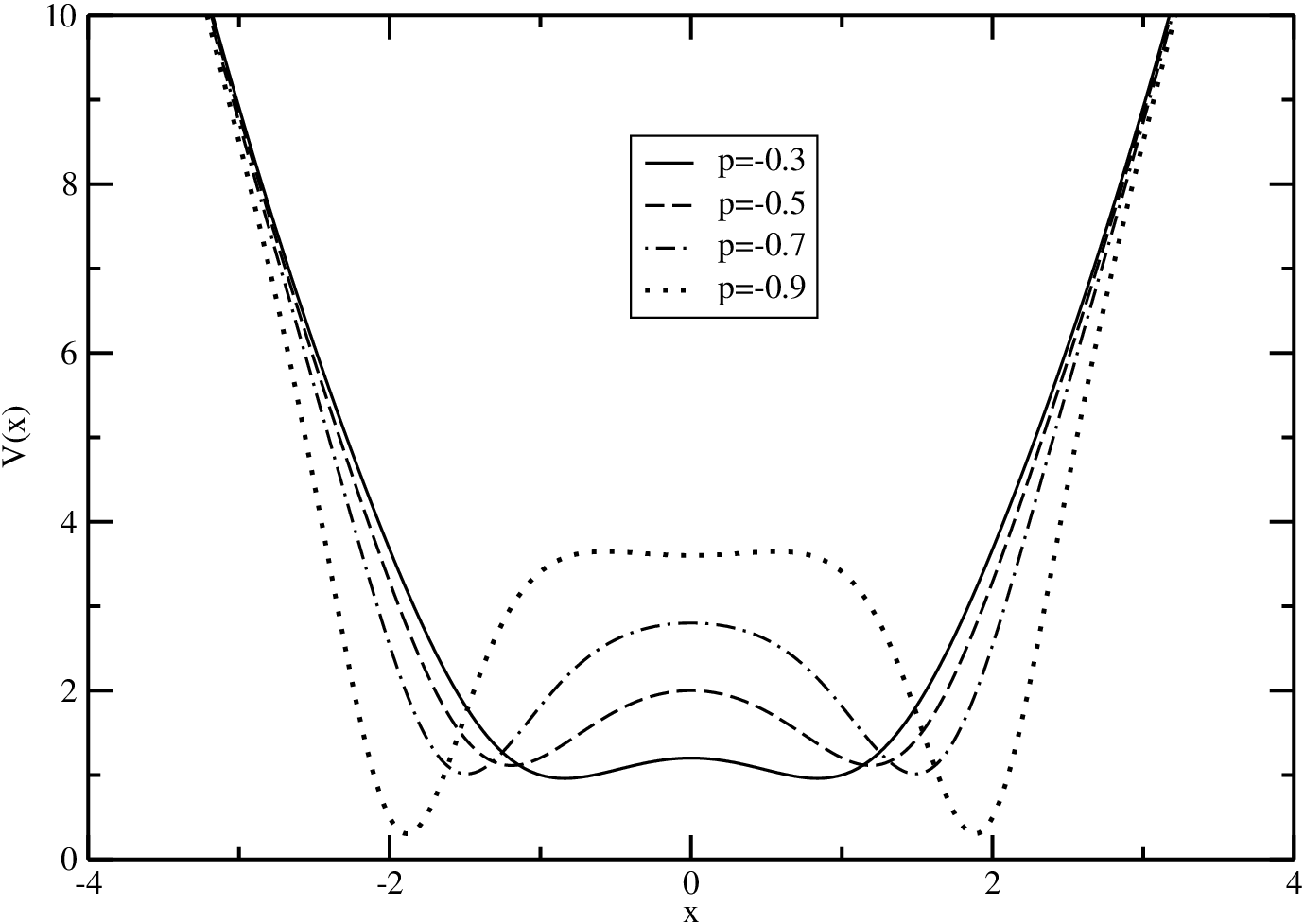}&
\includegraphics[width=3in]{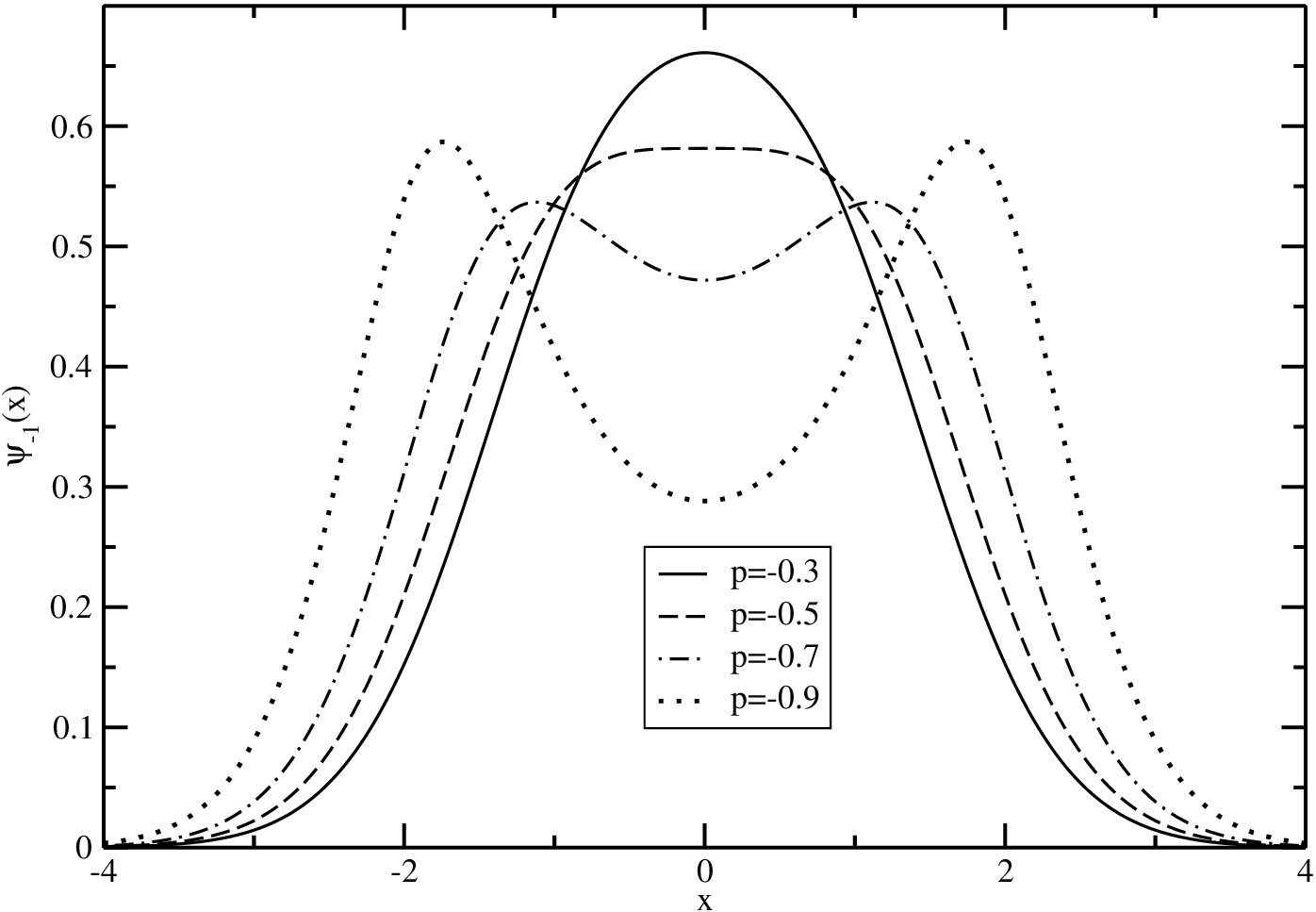}\\
\\
\\
\includegraphics[width=3in]{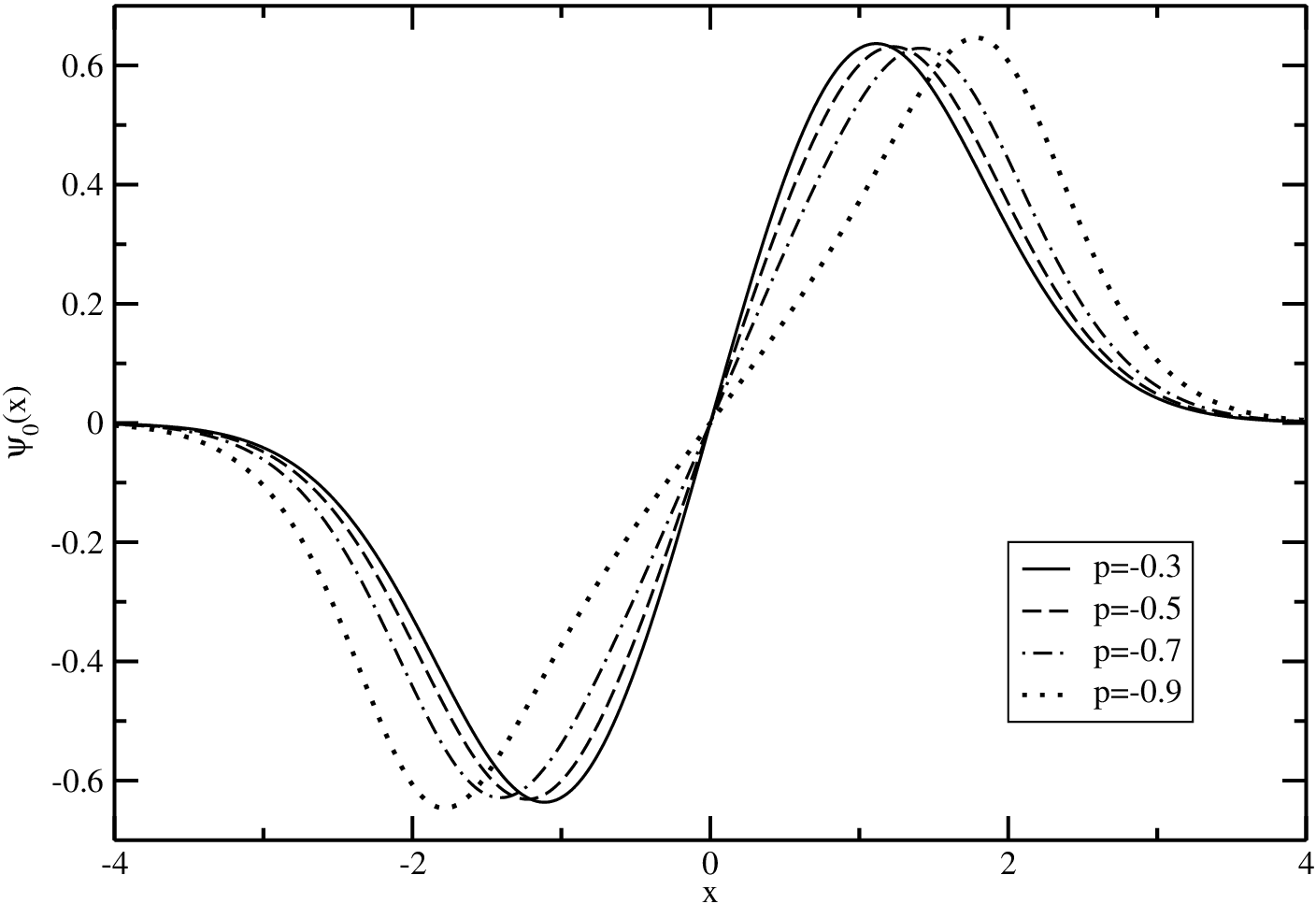}&
\ \ \includegraphics[width=3in]{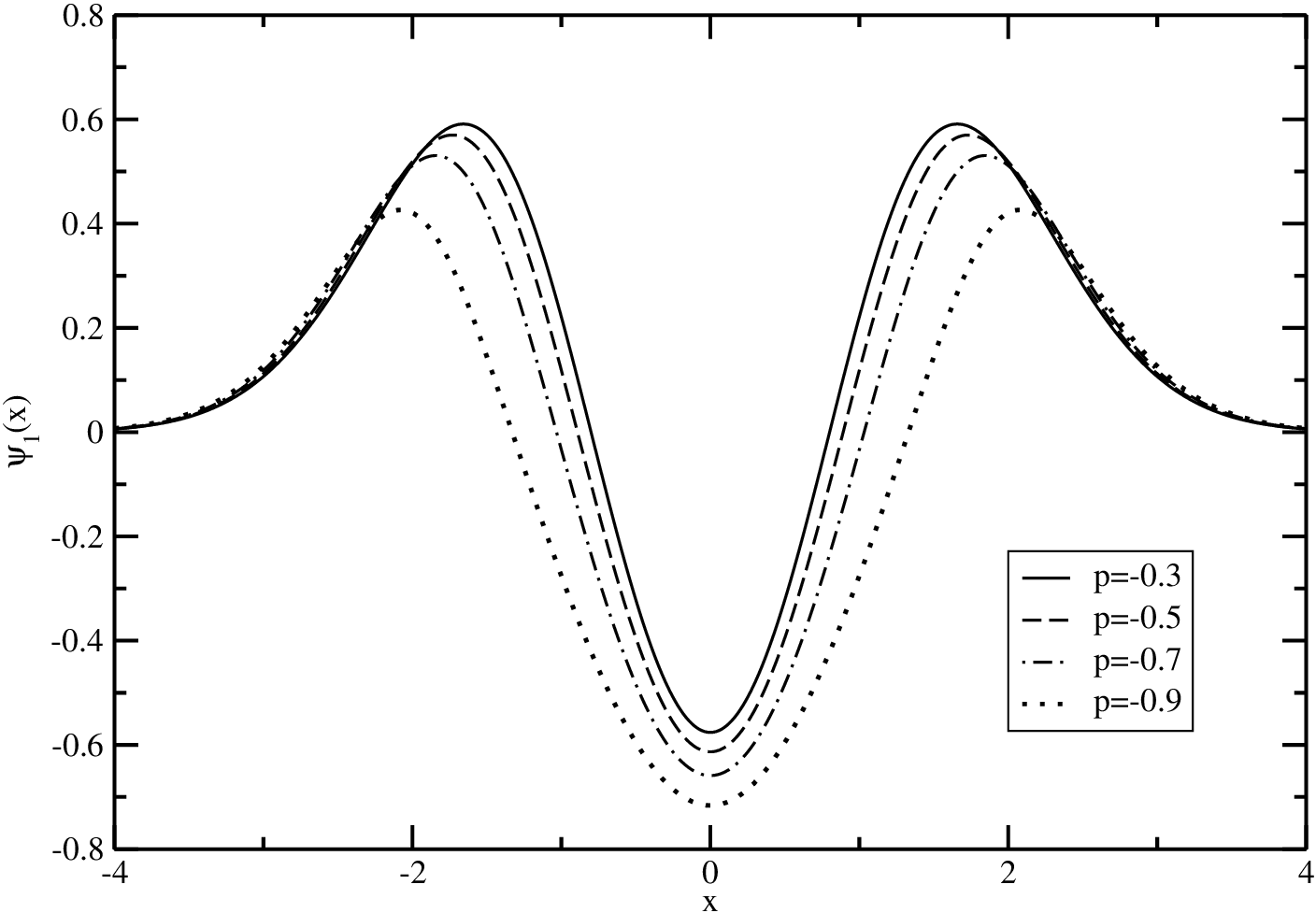}\\
\end{tabular}
\caption{Symmetric double well potentials and their wavefunctions corresponding to parameter values
$0>p>-1$ and $s=0$: (a) Potentials; (b) Ground states; (c) First excited states; (d) Second excited states.}
\end{figure}

If $0>p>-1$ the potential takes the form of a double well sitting at the bottom of a harmonic potential. 
The potentials and their first three wavefunctions are shown in Fig. (2) for $p=-0.3$, $-0.5$, $-0.7$ and
-$0.9$. The double well shape
can partly be understood as the effect of a ``negative dimple'' which pushes up the centre of the potential to
form a barrier between two wells. As $p$ approaches $-1$ the centre of the potential becomes flatter and then
develops a minimum which resembles the original parabolic minimum, but raised up by $4$ units of energy, whilst 
the two wells move up the sides of the harmonic potential. The ground state starts off with a single maximum
which then flattens out and becomes a minimum between two maxima. The wavefunction initially straddles the two
wells when they are relatively shallow, but as they become deeper and further apart, the wavefunction splits
into two pieces centred on the two wells. The first excited state wavefunction also initially straddles both wells
when they are shallow, and splits into two pieces as they become deeper. In the extreme limit where $p$ approaches 
$-1$, the ground state and first excited state take the form of symmetric and antisymmetric combinations of two
peaks centred on the two wells. This is to be expected since the lowest two energy levels are now very close together.
The second and higher excited states are relatively unaffected by the double well potential, and their wavefunction
always straddle both wells. The lowest two energy levels effectively form a two-level system with a large energy
gap to the remaining harmonic ladder.

\subsection{Skewed dimple potentials: $p>0$, $1>s>-1$}

\begin{figure}[h]
\begin{tabular}{cc}
\includegraphics[width=3in]{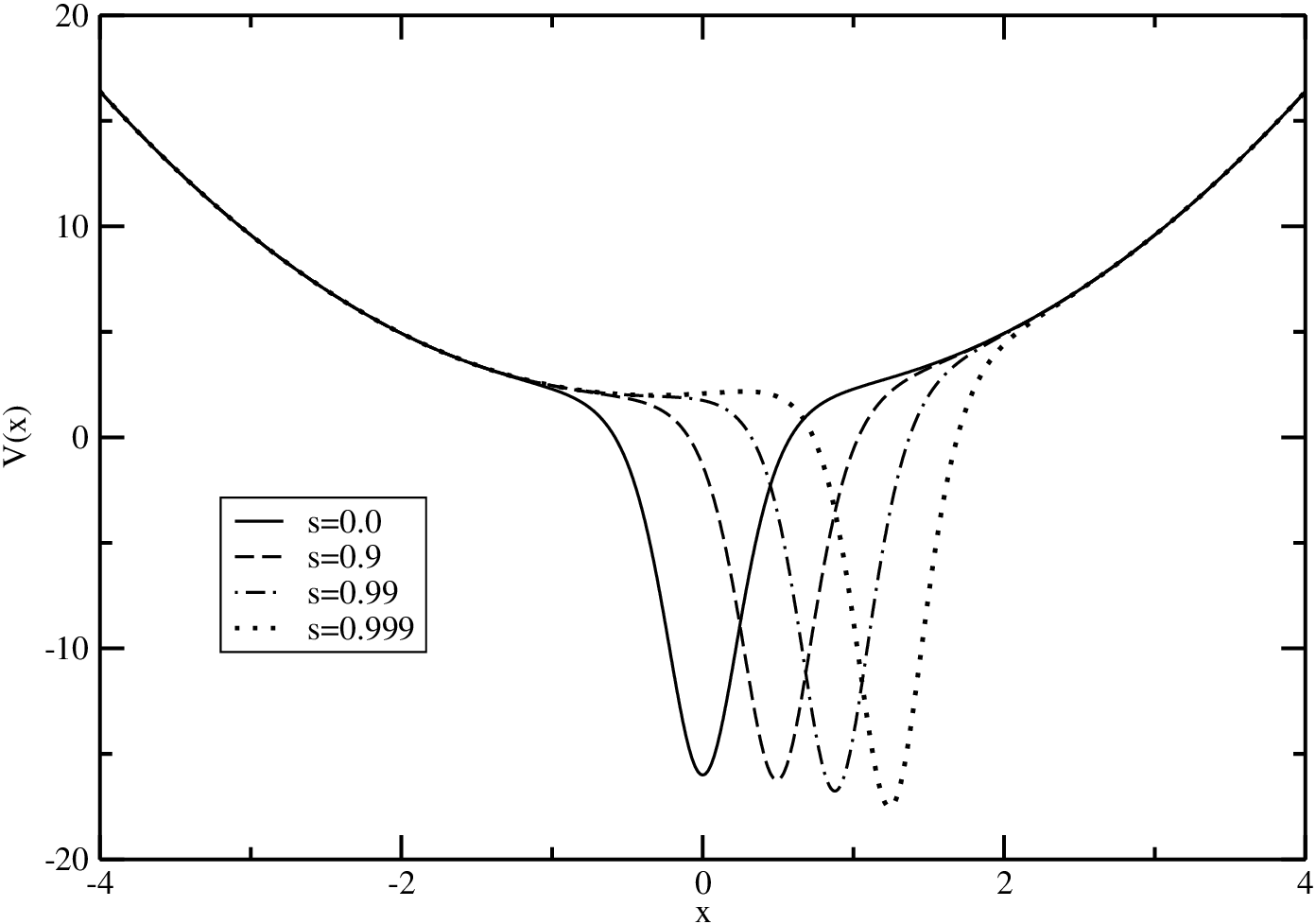}&
\includegraphics[width=3in]{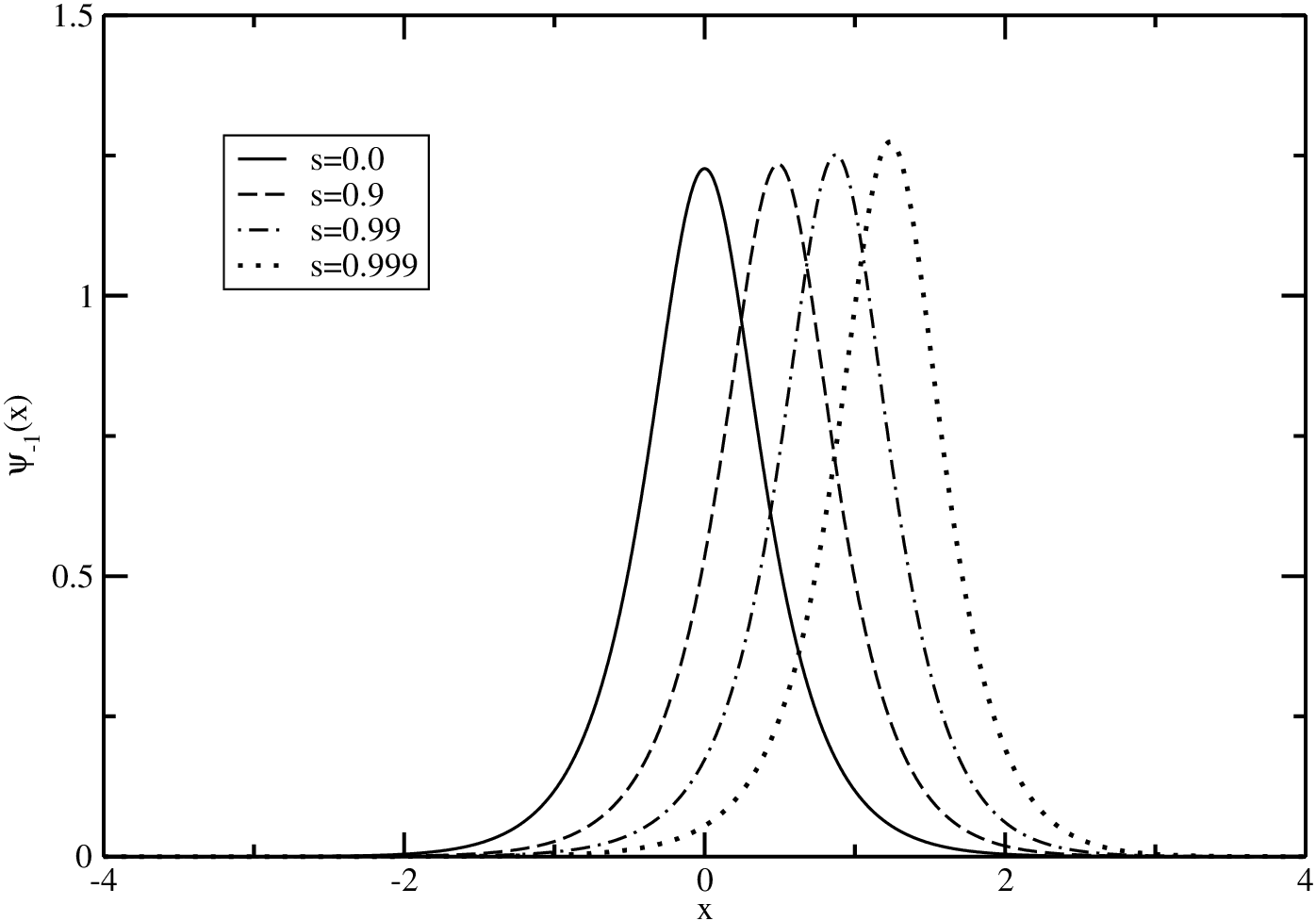}\\
\\
\\
\includegraphics[width=3in]{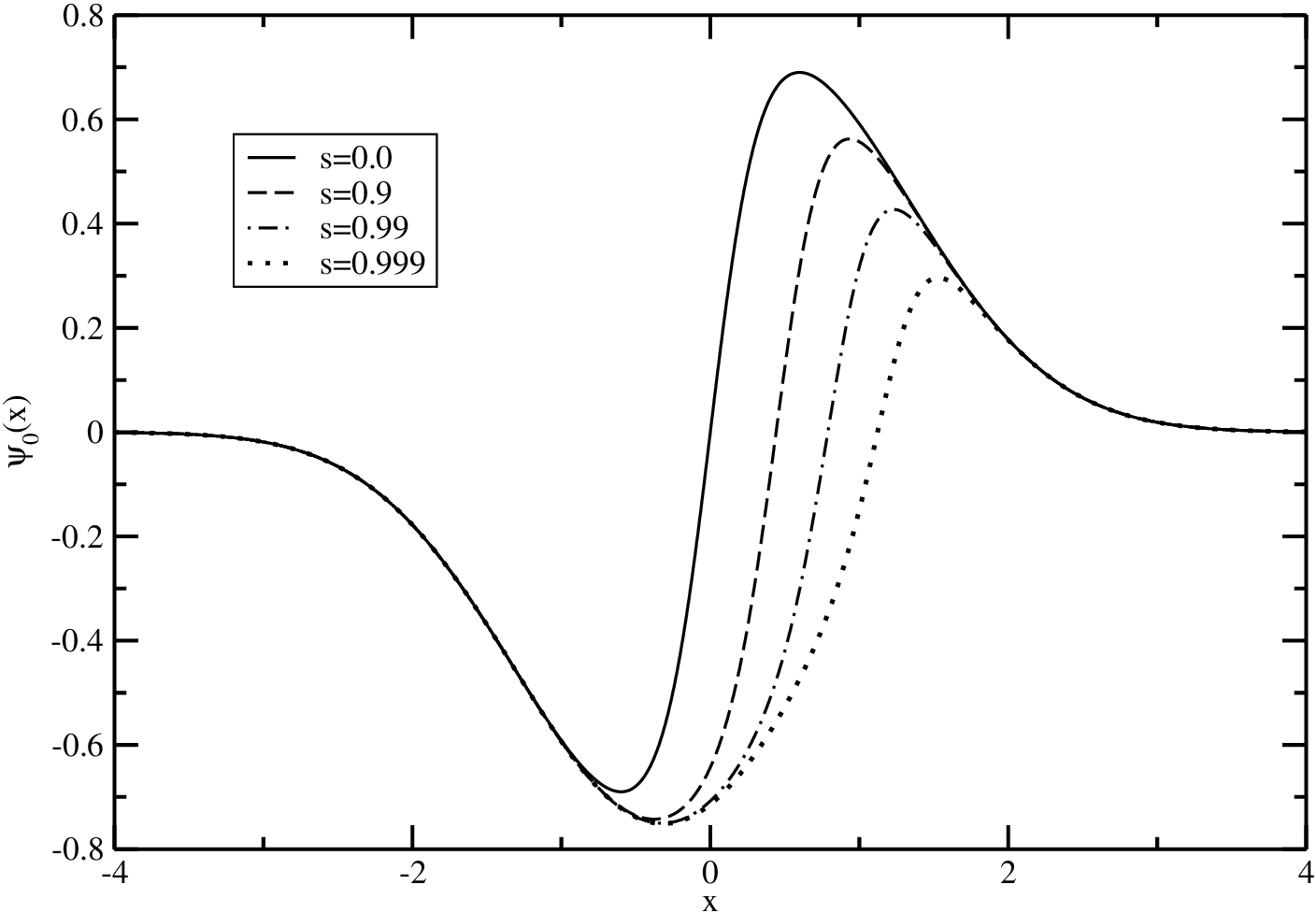}&
\ \ \includegraphics[width=3in]{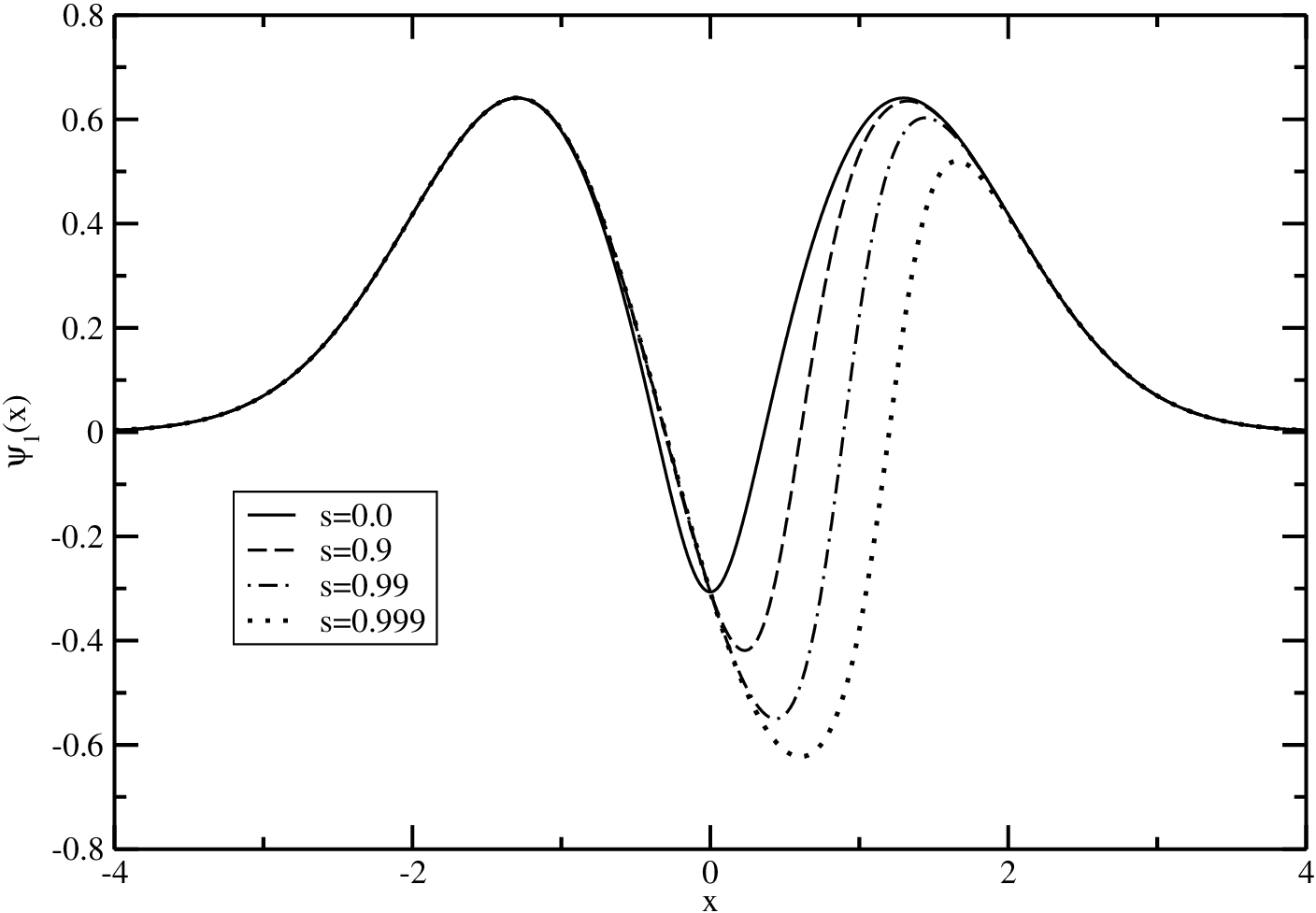}\\
\end{tabular}
\caption{Skewed dimple potentials and their wavefunctions corresponding to parameter values
$p=4.0$ and $1>s\ge 0$: (a) Potentials; (b) Ground states; (c) First excited states; (d) Second excited states.}
\end{figure}

If $p>0$ and $1>s>0$, the potential takes the form of an attractive dimple whose minimum is skewed to the right
(for $0>s>-1$ the mirror image is obtained). The potentials and their first three wavefunctions are shown in Fig. (3)
for $p=4$ and $s=0,0.9,0.99$ and $0.999$. The amount by which the dimple moves to the right depends
logarithmically on $1-s$. The ground state moves with the minimum of the dimple, and roughly maintains its gaussian
shape. The other excited states do not move with the dimple, and change their shape to retain orthogonality to the
ground state. The first excited state starts off with odd parity when $s=0$, and tends asymptotically to even parity
as $s$ approaches $1$ and the dimple moves off to infinity. This is because the orthogonality to the ground state becomes
less restrictive as the ground state wavefunction moves off to infinity with the dimple. The second excited state similarly
evolves from even parity at $s=0$ to odd parity as $s$ approaches $1$.

\subsection{Skewed double well potentials: $0>p>-1$, $1>s>-1$}

\begin{figure}[h]
\begin{tabular}{cc}
\includegraphics[width=3in]{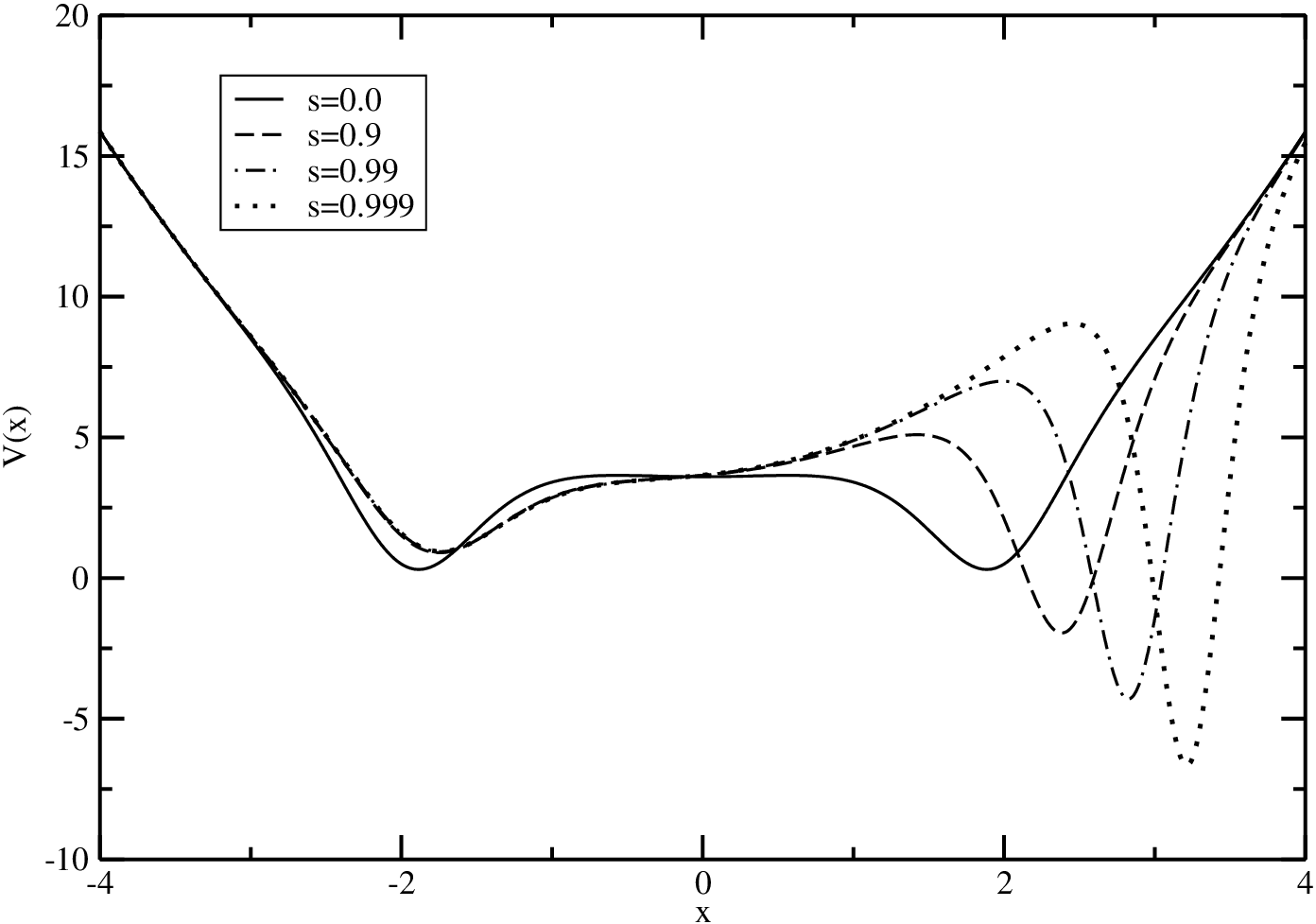}&
\includegraphics[width=3in]{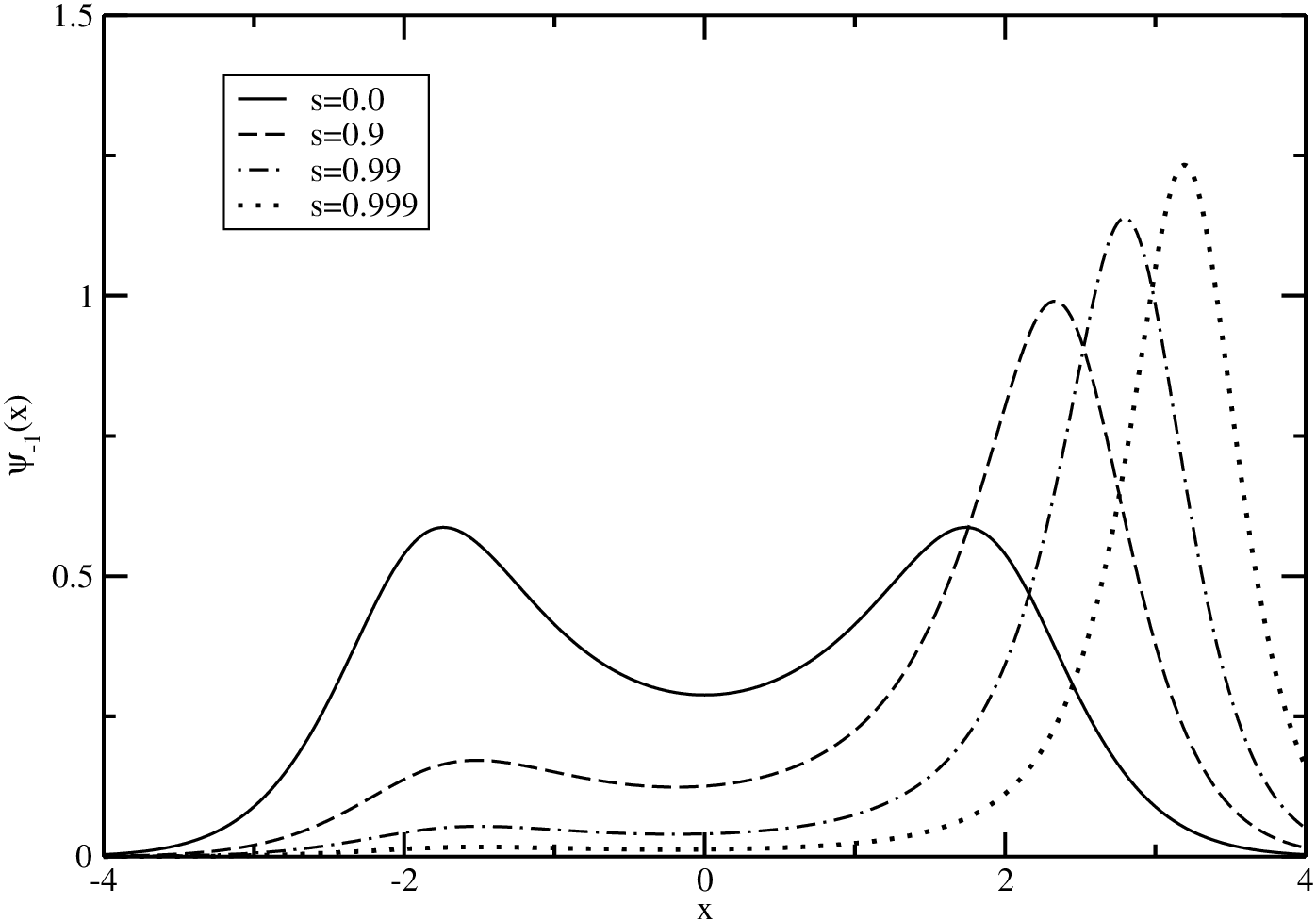}\\
\\
\\
\includegraphics[width=3in]{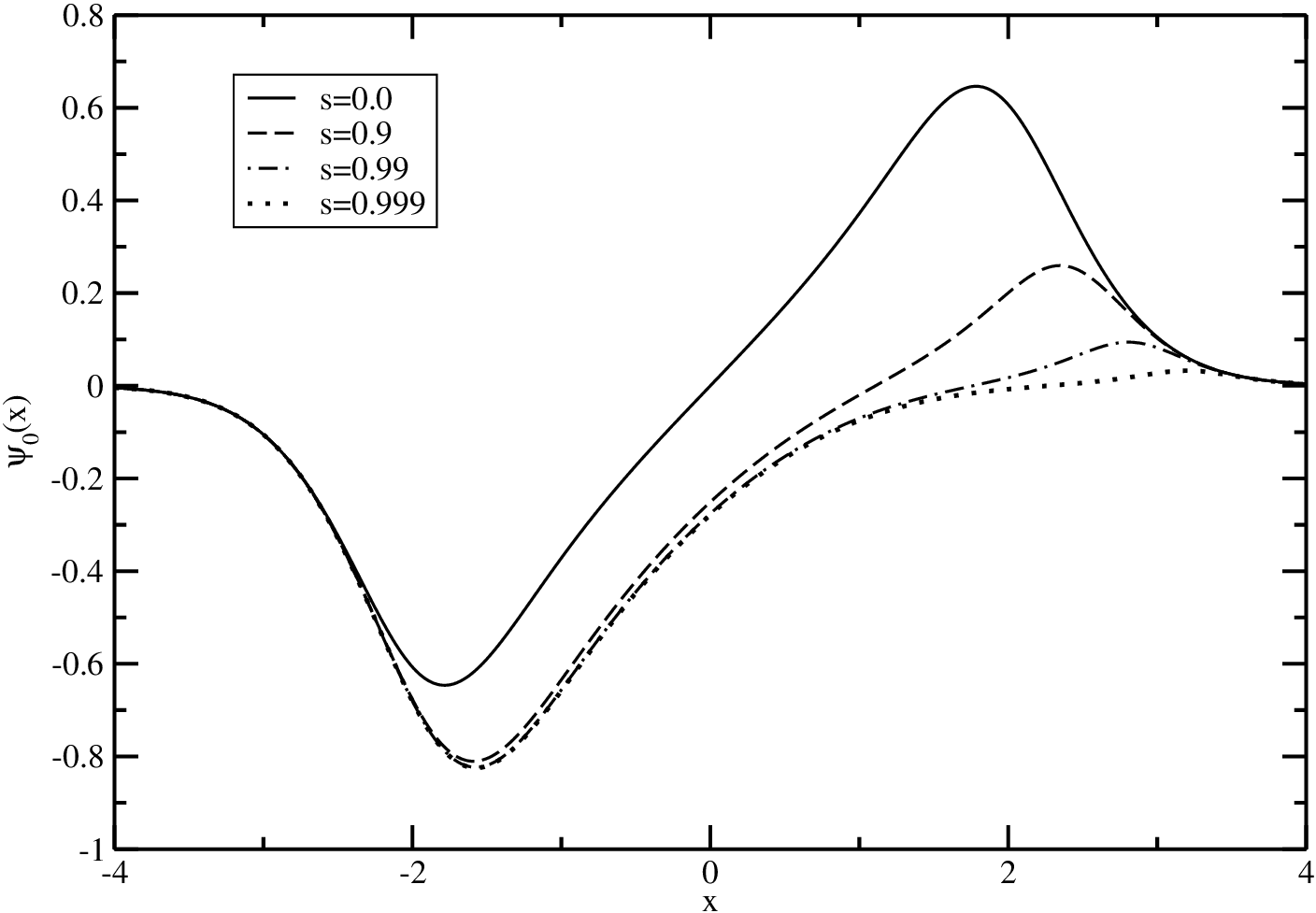}&
\ \ \includegraphics[width=3in]{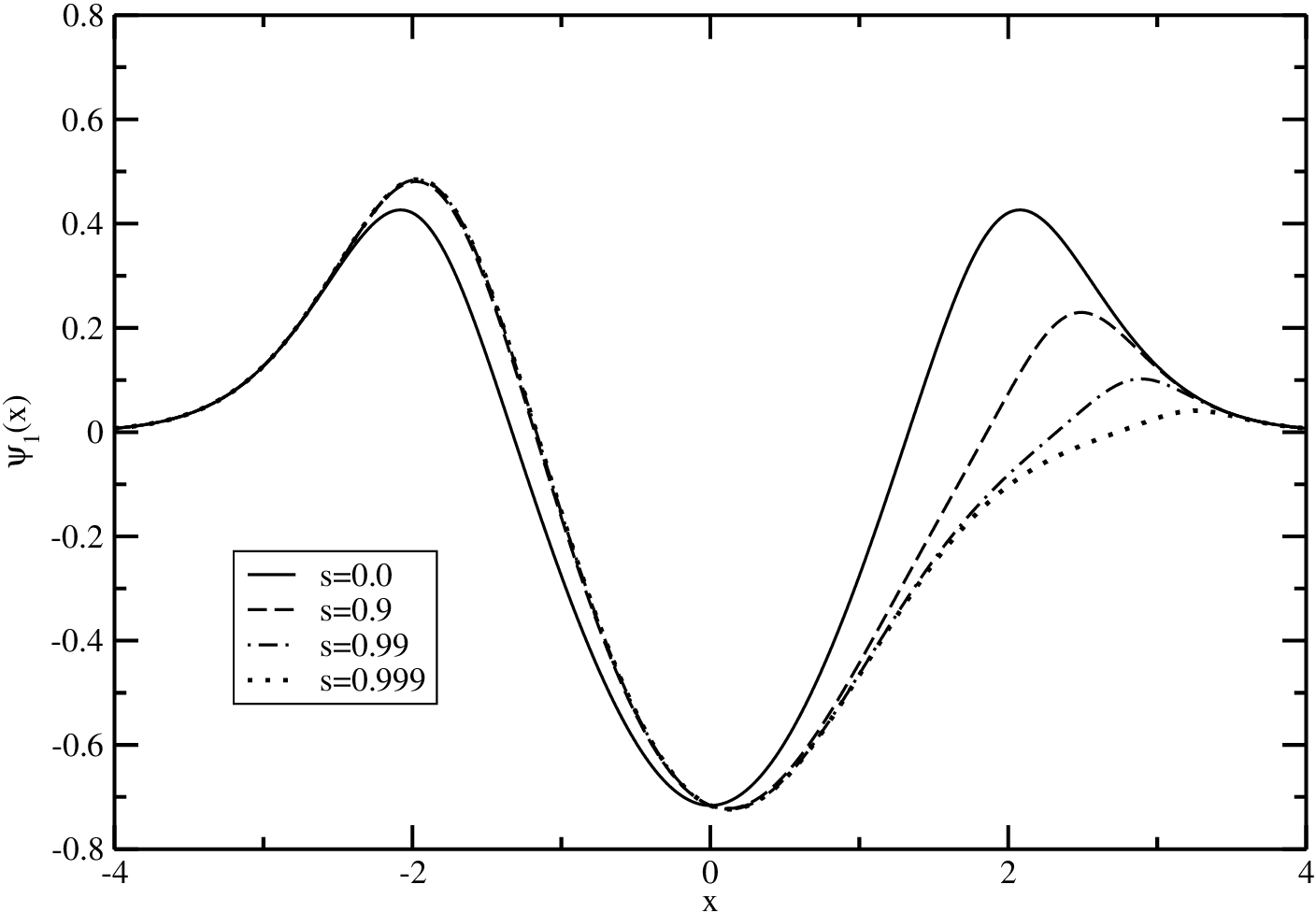}\\
\end{tabular}
\caption{Skewed double well potentials and their wavefunctions corresponding to parameter values
$p=-0.9$ and $1>s\ge 0$: (a) Potentials; (b) Ground states; (c) First excited states; (d) Second excited states.}
\end{figure}

If $0>p>-1$ and $1>s>0$, the potential takes the form of a double well with the right hand well deeper and further
from the origin than the left (for $0>s>-1$ the mirror image is obtained). The potentials and their first three
wavefunctions are shown in Fig. (4) for $p=-0.9$ and $s=0,0.9,0.99$ and $0.999$. The amount by which the deeper well
moves to the right, and its depth, depend logarithmically on $1-s$. In contrast the shallow well rapidly approaches
a limiting form. The ground state wavefunction has two peaks which are centred on the two wells. As $s$ increases, the
height of the peak in the right hand well increases, whilst that in the left decreases. As $s$ approaches $1$, the ground
state looks like a gaussian centred on the right well. The first excited state starts off with odd parity at $s=0$, with
equal sized peaks in two wells. As $s$ increases to $1$, the peak in the right hand well vanishes, and the wavefunction looks
like a skewed gaussian centred on the left hand well. It is as if the right hand well has moved so far away that the first excited
state is effectively the ground state of the left well. The second excited state starts off with even parity at $s=0$, and ends
up looking like the first excited state of the left well as $s$ approaches $1$. 

\section{BEC condensation temperature in a dimple trap}

Although Bose-Einstein condensation is not possible in one or two dimensions in a translationally invariant system, no such
restriction exists in a confined system \cite{Petrov2004}. Low-dimensional condensates are routinely produced in magnetic or optical traps that
are well modelled by a harmonic oscillator potential. The critical temperature, $T_c$, for a cloud of $N$ bosons in such a 
one-dimensional harmonic well can be
found from particle number conservation when the chemical potential equals the ground state energy. This gives the equation

\begin{displaymath}
N=\sum_{n=1}^{\infty} {1\over e^{n\beta_c\hbar\omega}-1}\approx 
{1\over\hbar\omega\beta_c}\int_{\hbar\omega\beta_c}^\infty {dx\over e^x-1}=
{1\over\hbar\omega\beta_c}\ln{\left({1\over 1-e^{-\hbar\omega\beta_c}}\right)},
\end{displaymath}
where $\beta_c=1/k_BT_c$. For large $N$ this approximates to
\begin{displaymath}
k_BT_c\approx {N\over\ln{N}}\,\hbar\omega.
\end{displaymath}
Experimentally the condensate fraction may be increased by adiabatically deforming the shape of the trapping potential 
\cite{Stamper1998}, so that
it takes the form of a harmonic well with an attractive dimple at the origin. Such dimple potentials are highly reminiscent of the
potentials discussed in section 4.1. These potentials allow
condensation to be achieved at a higher temperature, and in a reversible manner. Although heuristic semiclassical explanations of this 
phenomena exist in the literature \cite{PS2003}, it is preferable to have an exactly soluble quantum model. We will therefore consider Bose-Einstein
condensation in a dimple trap of the form discussed in section 4.1, with depth $4p(\hbar\omega/2)$. The energy levels of this potential are 
$E_0=(1-2p)\hbar\omega/2$ and $E_n=(2n+1)\hbar\omega/2$ ($n=1,2,3\dots$), so that the critical temperature is given by

\begin{displaymath}
\hspace{-1.0cm}
N=\sum_{n=1}^{\infty} {1\over e^{(n+p)\beta_c\hbar\omega}-1}\approx 
{1\over\hbar\omega\beta_c}\int_{\hbar\omega\beta_c}^\infty {dx\over e^{x+p\beta_c\hbar\omega}-1}=
{1\over\hbar\omega\beta_c}\ln{\left({1\over 1-e^{-(p+1)\hbar\omega\beta_c}}\right)}.
\end{displaymath}
The dependence of $T_c$ as a function of $p$ is plotted in Fig. (5) for a fixed number of atoms, $N=10^5$. As expected, $T_c$ increases monotonically with $p$,
and doubles when $p\approx 200$. This corresponds to a dimple depth of $4p(\hbar\omega/2)=400\hbar\omega$. This is easily achievable in experiments \cite{Garrett2011}, where
typically $\omega\approx 2\pi\times 10Hz$ so that $400\hbar\omega\approx 2nK$, whilst dimple depths of up to $1000nK$ may be produced.

\begin{figure}[t]
\begin{center}
\includegraphics[width=4in]{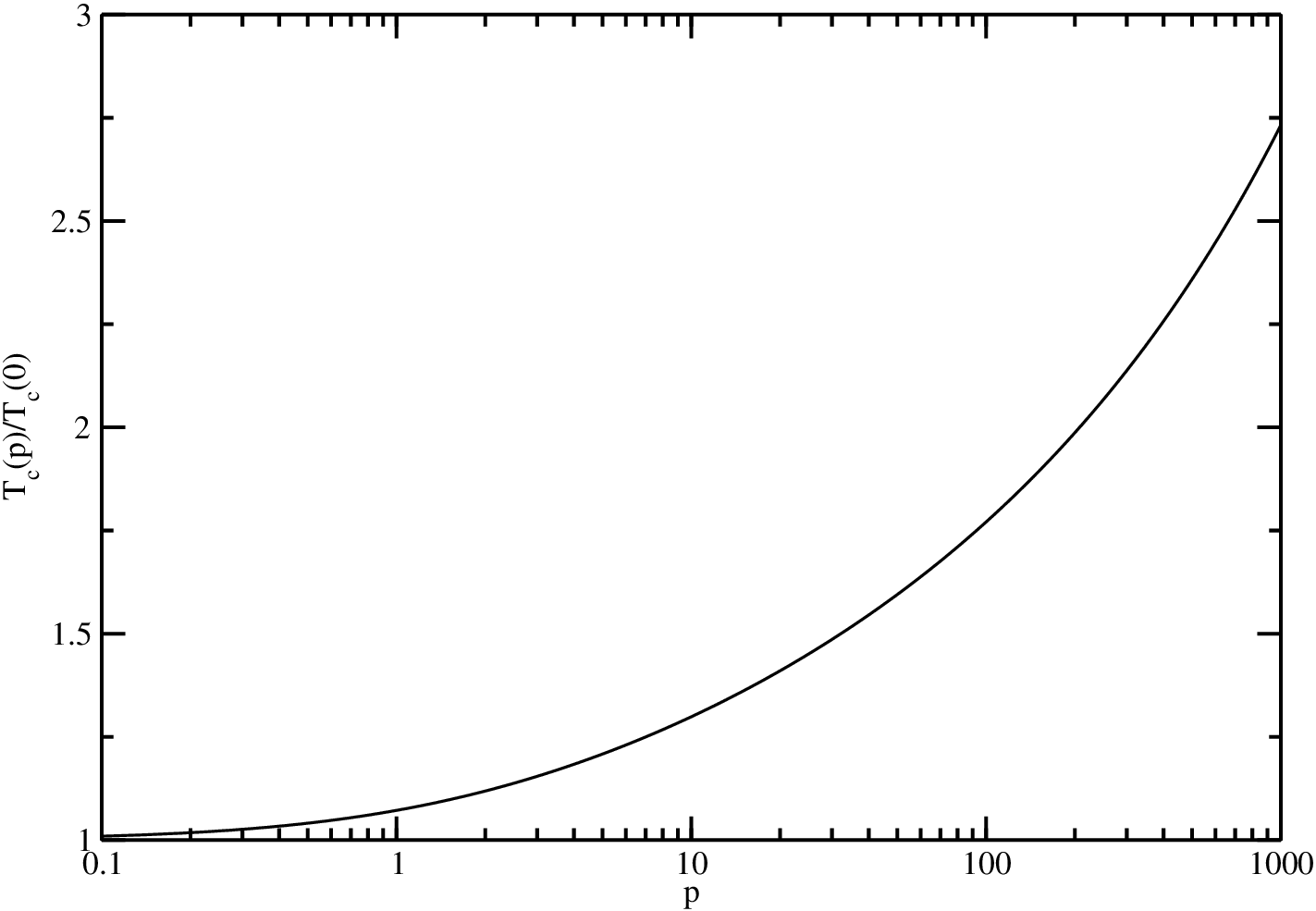}
\end{center}
\caption{Relative transition temperature, $T_c(p)/T_c(0)$, as a function of dimple depth parameter, $p$, for a cloud of $N=10^5$ bosons in a dimple trap.}
\end{figure}

\section{Summary and Conclusions}

In this paper we have derived the most general non-singular partner potentials of the quantum harmonic oscillator, and obtained their complete set of energy
eigenvalues and normalized eigenfunctions. These potentials depend upon two continuous parameters: a depth parameter, $p$, which tunes the depth of the centre
of the potential, and a skew parameter, $s$, which breaks the reflection symmetry of the potential. The spectra of these potentials are identical to that of
the quantum harmonic oscillator, except for the ground state whose energy is lowered by $2p$. A possible physical application of these potentials is to the
investigation of Bose-Einstein condensation in dimple traps.  

\appendix

\section{Normalization of ground state eigenfunctions}
Suppose that $\phi_{p,s_1}(x)$ and $\phi_{p,s_2}(x)$ are independent solutions of the Schr\"odinger differential equation,
\begin{equation*}
-{d^2\phi\over dx^2}+V(x)\phi=\epsilon_p\phi.
\end{equation*}
It follows that the Wronskian,
\begin{equation*}
W(x)=\phi_{p,s_1}(x)\phi'_{p,s_2}(x)-\phi'_{p,s_1}(x)\phi_{p,s_2}(x),
\end{equation*}
is a constant, which we will denote by $C(p,s_1,s_2)$. Dividing the Wronskian
by $\phi_{p,s_1}^2$ gives
\begin{equation*}
{C(p,s_1,s_2)\over\phi_{p,s_1}(x)^2}={\phi'_{p,s_2}(x)\over\phi_{p,s_1}(x)}
-{\phi'_{p,s_1}(x)\phi_{p,s_2}(x)\over\phi_{p,s_1}(x)^2}={d\over dx}\left[{\phi_{p,s_2}(x)\over\phi_{p,s_1}(x)}\right],
\end{equation*}
so that upon integration we get
\begin{equation*}
\int_{-\infty}^{\infty} {dx\over\phi_{p,s_1}(x)^2}=
{1\over C(p,s_1,s_2)}\left[{\phi_{p,s_2}(x)\over\phi_{p,s_1}(x)}\right]_{-\infty}^{\infty}.
\end{equation*}
For our problem
\begin{equation*}
\phi_{p,s}(x)=\left[\Phi({\textstyle{p+1\over 2}},{\textstyle{1\over 2}};x^2)
+sx\Phi({\textstyle{p\over 2}+1},{\textstyle{3\over 2}};x^2)\right]e^{-x^2/2},
\end{equation*}
which has asymptotic limit as $x\rightarrow\pm\infty$
\begin{equation*}
\phi_{p,s}(x)\sim\left[{\Gamma({\textstyle{1\over 2}})\over\Gamma({\textstyle{p+1\over 2}})}
+s{\Gamma({\textstyle{3\over 2}})\over\Gamma({\textstyle{p\over 2}+1})}\hbox{sgn}(x)\right]
|x|^p e^{x^2/2}
\end{equation*}
Evaluating the Wronskian at $x=0$ gives $C(p,s_1,s_2)=s_2-s_1$, so that the integral becomes
\begin{equation*}
\int_{-\infty}^{\infty} {dx\over\phi_{p,s_1}(x)^2}={1\over s_2-s_1}
\left[{\alpha+\beta s_2\over \alpha+\beta s_1}-{\alpha-\beta s_2\over \alpha-\beta s_1}\right]
={2\alpha\beta\over\alpha^2-\beta^2 s_1^2},
\end{equation*}
where for convenience we have defined $\alpha=\Gamma({\textstyle{1\over 2}})/\Gamma({\textstyle{p+1\over 2}})$
and $\beta=\Gamma({\textstyle{3\over 2}})/\Gamma({\textstyle{p\over 2}+1})$. It follows that
\begin{equation*}
\int_{-\infty}^{\infty} {dx\over\phi_{p,s}(x)^2}={\Gamma({\textstyle{p\over 2}+1})\Gamma({\textstyle{p+1\over 2}})\over
\Gamma({\textstyle{p\over 2}+1})^2-{s^2\over 4}\Gamma({\textstyle{p+1\over 2}})^2}=
{\sqrt{\pi}\,\Gamma(p+1)\over 2^p\left[\Gamma({\textstyle{p\over 2}+1})^2-{s^2\over 4}\Gamma({\textstyle{p+1\over 2}})^2\right]},
\end{equation*}
where in the last step we used the duplication formula $2^{2z-1}\Gamma(z)\Gamma(z+{\textstyle{1\over 2}})=\sqrt{\pi}\Gamma(2z)$
with $z=(p+1)/2$. We see that the simple formula obtained by guesswork based on analytic continuation from even integer $p$,
and the divergence of the integral at $s=\pm s_{max}$, is correct. Moreover, the Wronskian technique demonstrated above allows
one to normalize any new ground state wavefunction generated by the factorization method.

\end{document}